\documentclass[aps,twocolumn,floats,showpacs,prb]{revtex4}
\usepackage{amsmath,amssymb,graphicx,epsf,epstopdf,epsfig}
\newcommand{\etal}{{\it et al.}}

\begin{document}

\title{Lifshitz Transition in Underdoped Cuprates} 

\author{M. R. Norman}
\affiliation{Materials Science Division, Argonne National Laboratory, Argonne IL 60439}

\author{Jie Lin}
\affiliation{Materials Science Division, Argonne National Laboratory, Argonne IL 60439}

\author{A. J. Millis}
\affiliation{Department of Physics, Columbia University, 538 W 120th St., New York, NY 10027}

\begin{abstract}
Recent studies show that quantum oscillations thought to be associated with a density wave reconstructed Fermi surface disappear at a critical value of the doping for YBa$_2$Cu$_3$O$_{6+y}$, and the cyclotron mass diverges as the critical value is approached from the high doping side. 
We argue that the phenomenon is due to a Lifshitz transition where the pockets giving rise to the  quantum oscillations connect to form an open (quasi-1d) Fermi surface. The estimated critical doping is close to that found by experiment, and the theory predicts a logarithmic divergence of the cyclotron mass 
with a coefficient comparable to that observed in experiment.
\end{abstract}
\date{\today}
\pacs{74.25.Jb, 72.15.Gd, 75.30.Fv}

\maketitle

The unusual doping dependence of the physical properties of the high $T_c$ cuprate superconductors  has been a focus of interest since the beginning of the field. Experiments conducted over the past several  years have demonstrated that a crucial aspect of the doping dependence is a Fermi surface reconstruction, 
most likely due to some form of spin or charge density wave order. In overdoped materials,  photoemission \cite{RMP} and quantum oscillation studies \cite{Tl2201} have confirmed the existence of  a large Fermi surface, of size and shape compatible with band theory.  As the doping is reduced,
the form of the Fermi surface changes. Photoemission data indicate that the Fermi surface breaks up. The initial studies were interpreted in terms of  disconnected `Fermi arcs', \cite{Nat98} but some recent  studies have argued that what is observed is actually part of a closed hole pocket.\cite{PJ,Zhou}  One issue is that photoemission 
experiments access the `normal' (non-superconducing) state by raising the temperature above the superconducting transition temperature $T_c$. Quantum oscillation measurements, on the other hand, are conducted at high magnetic fields which  suppress superconductivity, permitting (at least in principle) access to the low temperature `normal' state. As the doping is reduced, unambiguous signatures  of the formation of small Fermi pockets are  observed,\cite{Doiron,LeBoeuf} in particular a dominant oscillation frequency of  about  530 Tesla, corresponding to a pocket size about 1.9\% of the Brillouin zone. The fate of the pocket (or pockets) as the doping is reduced is the subject of intense current interest. 

In a very interesting recent experiment, Sebastian \etal~\cite{Sebastian-QCP} report that in underdoped  YBa$_2$Cu$_3$O$_{6+y}$, the
cyclotron mass of the  Fermi pocket corresponding to the dominant quantum oscillation frequency 
diverges  near  $y$ of 6.46.\cite{Sebastian-QCP} Below  this doping, the oscillation frequency is not seen.  The critical  doping is close to  the value at which  
high field transport indicates localization,\cite{Boebinger,Proust} and also near the doping where  inelastic neutron scattering studies have indicated a collapse of the spin gap \cite{Dai} with the subsequent appearance of a nematic phase.\cite{Hinkov,Dai2}  Ref.~\onlinecite{Sebastian-PNAS} suggests that the mass divergence  might be associated with an excitonic instability involving electron  and hole pockets. 

\begin{figure}[htbp]
\centerline{\includegraphics[width=0.85\columnwidth]{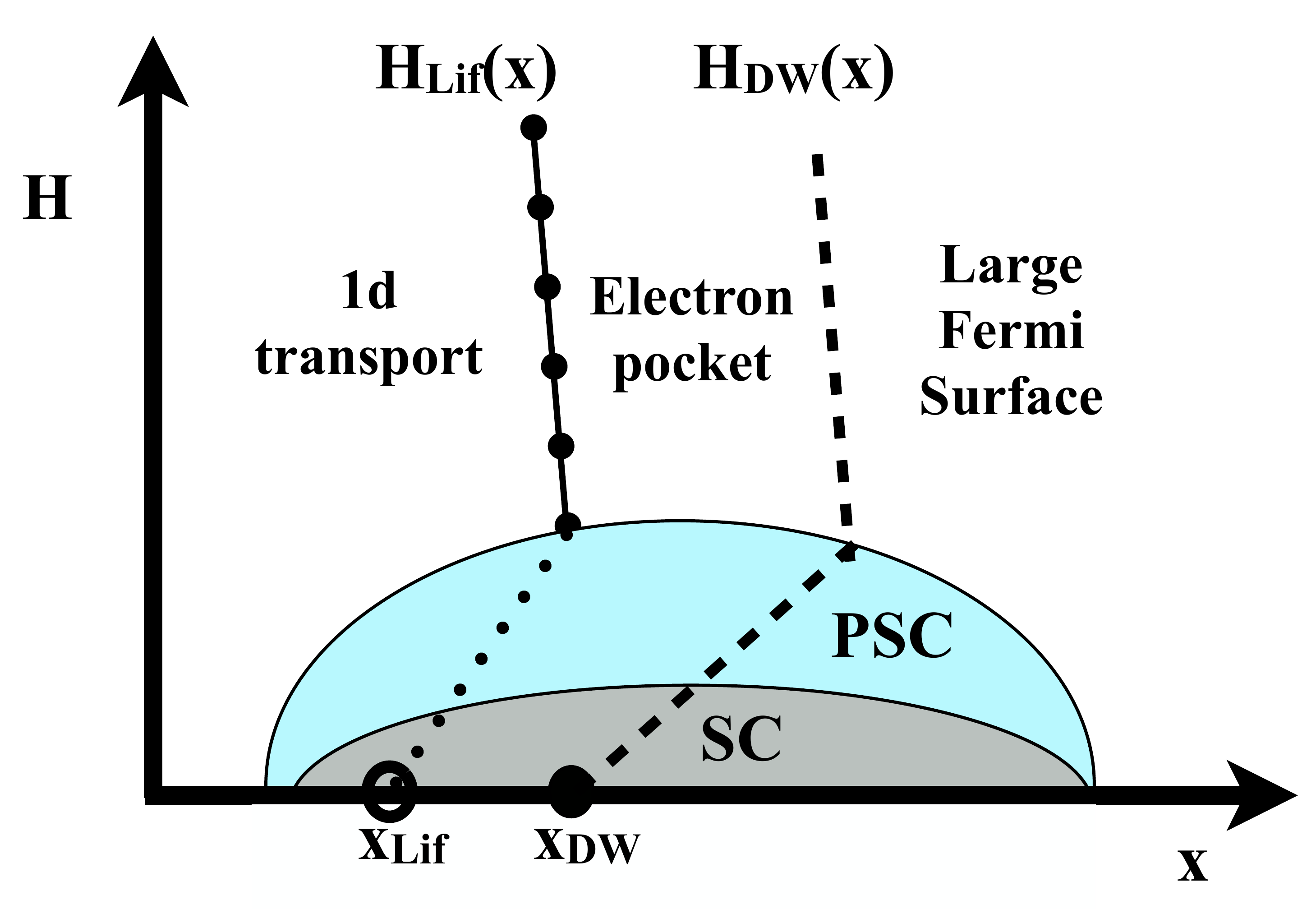}}
\caption{(Color online) Schematic phase diagram at zero temperature in the plane of doping $x$ and applied magnetic field $H$ 
suggested by quantum oscillation experiments and the present work.  Shaded regions: superconducting (zero resistance) phase ($SC$) and  regime of precursor (fluctuating) superconductivity ($PSC$). $H_{DW}(x)$ (dashed line): onset of density wave order (broken lattice translation  symmetry). At fields above the $PSC$ boundary, the Fermi surface is well defined and, in the presence  of density wave order, is reconstructed as indicated in the figure.  $H_{Lif}(x)$ (solid line with filled circles): Lifshitz transition proposed in the present work, at which the dominant Fermi pockets  connect to form an open Fermi surface.   Light dotted line  indicates the continuation of the Lifshitz transition into the $PSC/SC$ regimes, where the gapping of the Fermi surface converts it to a crossover.}
\label{fig1}
\end{figure}

In this paper we propose that the transition observed by Sebastian \etal~\cite{Sebastian-QCP} is a   Lifshitz transition where the pockets touch and so connect to form an open (quasi-1d) Fermi surface.   The phase diagram resulting from our proposal  is shown in Fig.~\ref{fig1}.
The new feature added to existing phase diagrams such as those of Ref.~\onlinecite{Subir} is the Lifshitz transition, shown as the heavy line with filled circles. We present theoretical  calculations showing that for reasonable parameters, such a transition can occur at the experimentally  observed doping.  In two dimensions, we find that the cyclotron mass diverges logarithmically  at the Lifshitz transition. We estimate the doping dependence of the mass
and the magnitude of the divergence, finding good  agreement with the data.     
The sensitivity of one dimensional conductors to  localization \cite{LR} is consistent with the divergent resistivity observed at lower dopings.\cite{Boebinger,Proust}  Our proposal thus naturally explains the main features of the observations near this doping. 

\begin{figure}[b]
\centerline{\includegraphics[width=3.4in]{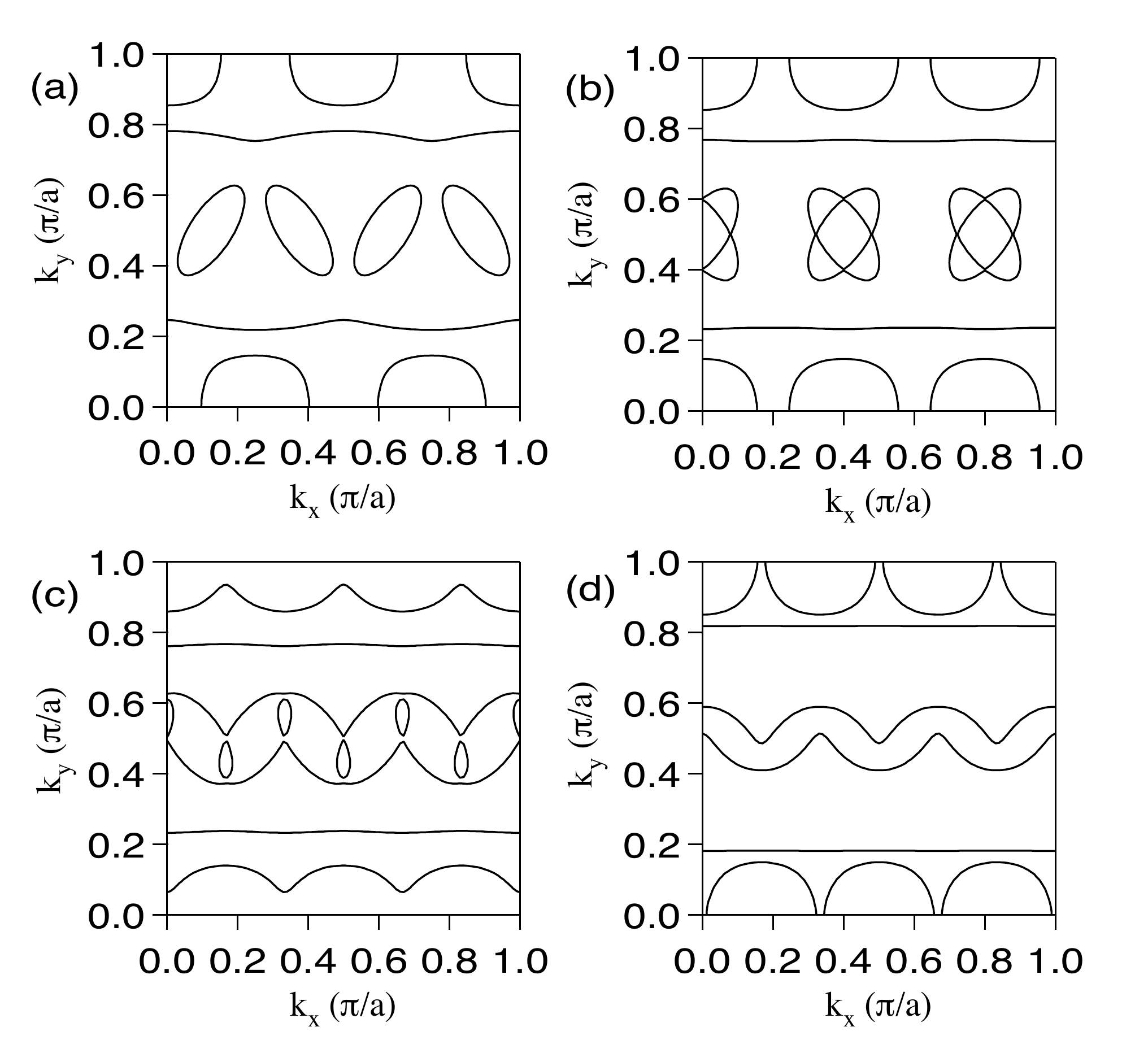}}
\caption{Fermi surface for (a) $V_s$=0.178 eV, $x=\delta$=1/8,
(b) $V_s$=0.204 eV, $x=\delta$=1/10, (c) $V_s$=0.233 eV, $x=\delta$=1/12,
and (d) $V_s$=0.25 eV, $V_c$=-0.12 eV, $x=\delta$=1/12,
where $V_s$ is the spin potential, $V_c$ the charge potential, $x$ the doping, 
and $\delta$ the incommensurability.}
\label{fig2}
\end{figure}

Our  calculations are based on a linear  spin density wave (magnetic stripe)  model introduced to  account for the first generation of quantum oscillation experiments.\cite{Millis} The model involves  electrons described by a tight binding band structure believed to be appropriate for hole-doped  high $T_c$ superconductors and subject to a periodic potential appropriate for an antiphase spin density  wave state characterized by the wavevector $q=(1-2\delta,1)\pi$, which we measure in units of 1/$a$ where $a$ is the lattice constant. Details of the electronic dispersion used and the form of the secular matrix can be  found in Ref.~\onlinecite{Millis}. In our previous work,\cite{Millis,Jie1} we focused on the case of  $\delta$=0.125. Electron pockets centered at $(0,\pi)$ and symmetry related points were found,  as well as hole pockets and open orbits. The precise fermiology depended on the specific model  parameters chosen, but the generic features of the calculation were the electron pockets and the  open orbits. The hole pockets were less robust in that they existed for smaller ranges of the density  wave potential.  We therefore argued that the observed quantum oscillation signal \cite{Doiron,LeBoeuf}  arose from the electron pocket. We will return to this issue below. 

In this paper we extend our analysis to lower dopings, $x < 0.125$. Neutron scattering  data \cite{Yamada} indicate that the stripe wavevector $\delta=x$, and we make this assumption in the calculations presented in this paper.  For simplicity, we include in 
most of our calculations only the  fundamental harmonic of the spin potential ($V_s$), but do show one example with a non-zero  second harmonic (charge) potential ($V_c$). Representative results are shown in the four panels of Fig.~\ref{fig2}.
Fig.~\ref{fig2}a reproduces our previous results for  $x$=1/8.\cite{Millis} The subsequent panels show the evolution of the Fermi surface as the doping  is reduced. In these calculations  $V_s$ was adjusted so  that the area of the electron pocket corresponds to an oscillation  frequency of $530$  Tesla as observed by experiment (experiment indicates only a weak doping dependence of the frequency \cite{Sebastian-QCP}). 
One sees that between $\delta=1/10$ and $\delta=1/12$, the pockets touch, resulting in a Lifshitz transition. The critical  $\delta$ can be easily estimated. In a repeated zone scheme, the pocket centers are separated  by a momentum 2$\delta$, so if $\delta$ is decreased while the pocket area is held fixed, the pockets 
must touch.  If we assume a circular pocket, which is consistent with a recent quantum oscillation study where the field angle was swept,\cite{Sebastian-PNAS} then for a pocket radius corresponding  to the oscillation frequency of  $530$ Tesla, and with the lattice constant, $a$, of 3.85 $\AA$, implies a critical value of $\delta$ equal to 0.078, corresponding to a period just beyond 12.  The parameters used to construct Fig.~\ref{fig2} lead to pockets slightly elongated along the $k_x$ direction, and the critical $\delta$ in our calculation is correspondingly slightly greater than $1/12$.  
The critical doping  can be  changed by introducing a second harmonic (charge) potential ($V_c$). Fig.~\ref{fig2}d shows that a negative value of $V_c$ makes the pocket shape more circular, stabilizing a closed pocket for the 12 period case (a positive $V_c$ would act oppositely by further elongating the  pocket along $k_x$).   

We identify the Lifshitz  transition at which the electron pockets vanish with the transition observed by Sebastian \etal~\cite{Sebastian-QCP}   This argument relies on the identification of  the $(0,\pi)$ pocket as the one which gives rise to the dominant $530$ Tesla quantum oscillation frequency. 
In the calculation this pocket is an electron pocket. The first quantum oscillation study \cite{Doiron}   of underdoped YBa$_2$Cu$_3$O$_{6+y}$ (YBCO) detected only  the dominant  oscillation frequency,  but interpreted it  as a hole pocket (as suggested also by the photoemission experiments). However, the 
subsequent observation of a negative Hall number in this doping range \cite{LeBoeuf} led to the  suggestion that the observed frequency originated from an electron pocket near the $(0,\pi)$ point of the  Brillouin zone.\cite{LeBoeuf,Millis} Recently, multiple frequencies have been seen.\cite{Audouard,Sebastian} The smaller ones, near the originally observed frequency, have been interpreted as arising from bilayer  splitting and  warping of the two dimensional Fermi cylinders associated with the $(0,\pi)$ pocket,\cite{Audouard} although a recent proposal suggests electron and hole cylinders of comparable size,\cite{Sebastian-PNAS} one of them warped and the other not.  

Further support for an electron pocket comes from 
the fact that a $\pi$ phase shift is observed between the  longitudinal and Hall Shubnikov-deHaas oscillations as expected for an electron pocket.\cite{Jaudet} Recently, it has been determined that the transport data are most consistent with a coherent  electron contribution, and an incoherent hole contribution.\cite{Rourke} As the doping is reduced,  the coherent electron contribution is lost.\cite{Louis} Also, the Shubnikov-deHaas oscillations associated with this oscillation frequency  are largest  for the c-axis resistance,\cite{Cyril} suggesting again an electron pocket since the c-axis hopping  is largest in the $(0,\pi)$ region of the zone.\cite{Sudip}

A difficulty with an interpretation of the dominant frequency in terms of electron pockets is that photoemission experiments at these dopings indicate a large  energy gap in this region of the Brillouin zone,\cite{RMP,Nat98,PJ,Zhou} suggesting that an electron pocket would not exist.
However, the photoemission and quantum oscillation experiments are not in direct contradiction. The  photoemission experiments are conducted at zero magnetic  field and at temperatures above $T_c$  (below $T_c$, one observes a d-wave superconducting gap which is maximal at the $(0,\pi)$  points). On the other hand, quantum oscillation experiments are conducted at high fields and at low temperatures. In that context, it is known that application of a magnetic field stabilizes the formation of the density wave state as observed by neutron scattering,\cite{Keimer} as indicated in Fig.~\ref{fig1}. Moreover, quantum oscillations can exist in the presence of an energy gap.  For instance, oscillations are observed in type II superconductors well below the upper critical field.\cite{Corcoran} For all of these reasons, we argue that the  main quantum oscillation frequency arises from the electron pocket located near the $(0,\pi)$ region  of the Brillouin zone, as found in our calculations.\cite{Millis}

\begin{figure}[b]
\centerline{\includegraphics[width=3.4in]{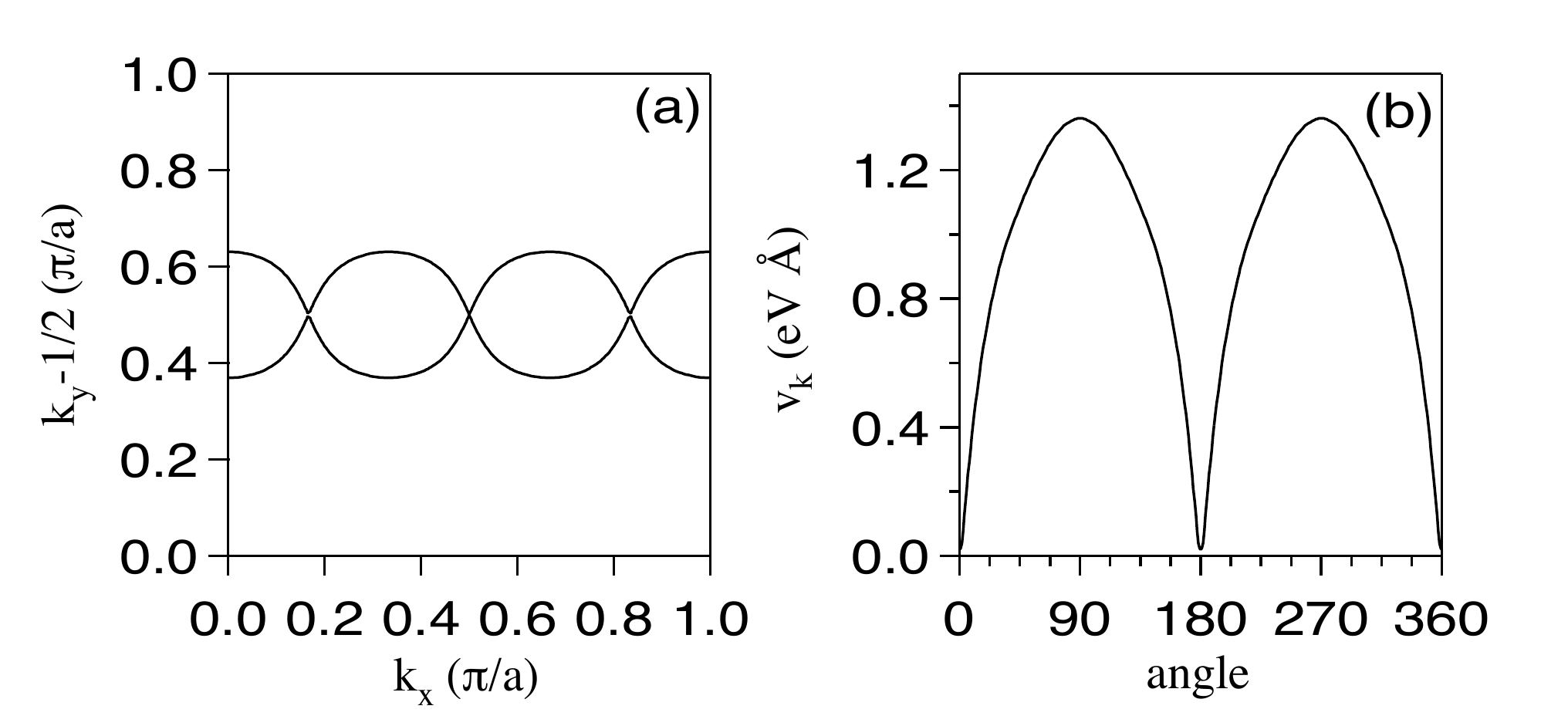}}
\caption{(a) Electron Fermi surface for $V_s$=0.235 eV, $\mu$=-0.35466 eV and $\delta$=1/12,
where $\mu$ is the chemical potential (for clarity, the rest of the bands are not shown).
(b) Fermi velocity around the Fermi surface from (a).
0 and 180 degrees correspond to the touching points.}
\label{fig3}
\end{figure}

\begin{figure}
\centerline{\includegraphics[width=3.4in]{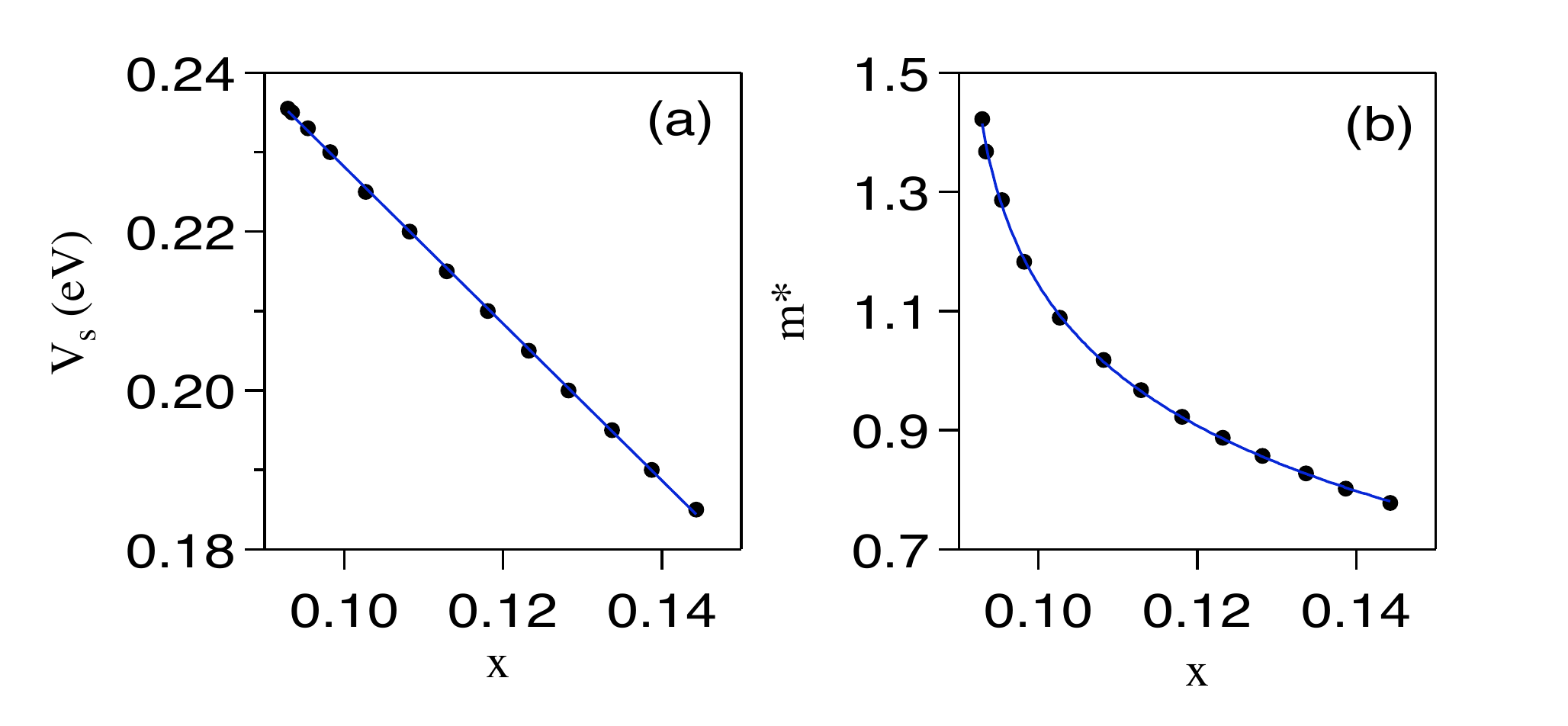}}
\caption{(Color online) (a) $V_s$ and (b) cyclotron mass versus doping, $x$, for  $\delta$=1/12. Note the actual cyclotron mass will be renormalized upwards by about a factor of three due to many body correlations not included in the present calculation. Points: results of calculations using the stripe SDW model. Curves: fits to linear (panel a) and logarithmic (panel b) functions.  $V_s(x)$ is determined by requiring that the area of the electron pocket  be doping independent. Obtaining an area corresponding to a quantum oscillation frequency of 530 Tesla at a critical doping of 9\% requires a nonzero $V_c$ as in Fig.~2d. To reduce the number of varying parameters, we set $V_c=0$ in this calculation; the $V_s$-only model for $\delta$=1/12 gives a frequency of 447 Tesla. }
\label{fig4}
\end{figure}

To study the Lifshitz transition  in more detail, we tune the system to the Lifshitz point and present in Fig.~\ref{fig3} the Fermi surface and the variation around the Fermi surface  of the Fermi velocity. One can see the linear variation of the Fermi contour about the touching point.  This is reflected in a linear variation of the Fermi velocity, $v_k$, about the touching point.  As a consequence, the cyclotron mass, which is defined by the line integral along the orbit, $\int dk/v_k$ (equivalent to the energy derivative of the cyclotron area), is logarithmically divergent.  This divergence can be seen analytically by noting that the touching point corresponds to a saddle point in the dispersion, which also leads to a logarithmically divergent density of states in two dimensions.  This divergence will be cut off by 
doping away from the Lifshitz point or by any c-axis warping of the Fermi cylinders, although we note that the oscillation studies have not been able to resolve the multiple frequencies expected from warping as the critical doping is approached. The divergence with doping is  illustrated in Fig.~\ref{fig4}b
(note that we plot the band mass; the actual mass will be renormalized by about a factor of three due to many body correlations as observed in photoemission).
In this plot, we take a simplified approach of working at fixed wavevector and fixed area of the cyclotron orbit.  This leads to a linear variation of $V_s$ with $x$, as illustrated in Fig.~\ref{fig4}a.
If instead $V_s$ was held fixed to its value at the critical doping (0.235 eV), the pocket area would decrease by a factor of three over the doping range indicated.

A more detailed  connection with experiment requires a knowledge of the true dependence of the wavevector on doping.  In zero field neutron scattering data, it is difficult to determine what the actual $\delta$ is because of the presence of the spin gap.  Quoted values of $\delta$ are typically
obtained from the momentum dependence of the neutron scattering intensity at an energy near  the spin gap energy, but as the spin branch below the resonance energy has a significant dispersion, these quoted values are underestimates of the zero energy value. It is also possible (Fig.~\ref{fig1}) that $\delta$ will change in a magnetic field. The other issue is that the relation of $y$, $T_c$, and the doping, $x$, is complicated because of the presence of chains in the YBCO structure, with differences depending on whether the sample is stabilized in the ortho-I or ortho-II structure.\cite{Liang}  Measurements
made at $y$=6.45 where no spin gap is present \cite{Dai,Li} 
find $\delta$=0.054, which  would be a lower bound.  An upper bound would be near 0.1, the expected value for $y$=6.5 where quantum oscillations have been seen. This indicates a rapid change of $\delta$ between these values of $y$.  Our estimated $\delta_c$ and therefore $y_c$ is safely within this range.  The bounds could be tightened significantly from  neutron scattering data near $y_c$ if a magnetic field is applied to induce an elastic signal.\cite{Keimer} A more decisive test of this scenario would be to use field angle sweeps \cite{Sebastian-PNAS} to look for the expected deviation of the orbit cross section from circular behavior as the proposed Lifshitz transition is approached.  One could also look for whether the field dependence of the transport indicates the presence of open orbits for $y < y_c$.

To summarize, we have proposed a simple explanation for the disappearance of quantum oscillations near $y$=6.46 in YBCO as due to a Lifshitz transition of electron pockets.  For $y$ less than this value, we propose that the Fermi surface is quasi-one dimensional (Fig.~\ref{fig1}), and thus subject to localization.  This would then account for the well known metal-insulator transition observed in high magnetic fields near this doping value.\cite{Boebinger,Proust}

We thank Louis Taillefer and Cyril Proust for sharing with us their unpublished data, and Suchitra Sebastian and Neil Harrison for discussions. AJM was supported by NSF-DMR-0705847, and MRN and JL by the U.S. DOE, Office of Science, under contract DE-AC02-06CH11357.


\begin{thebibliography}{99}

\bibitem{RMP} A. Damascelli, Z. Hussain and Z.-X. Shen,
Rev. Mod. Phys. {\bf 75}, 473 (2003).

\bibitem{Tl2201} B. Vignolle, A. Carrington, R. A. Cooper, M. M. J. French, A. P. Mackenzie, 
C. Jaudet, D. Vignolles, C. Proust and N. E. Hussey, Nature {\bf 455}, 952 (2008).

\bibitem{Nat98} M. R. Norman, H. Ding, M. Randeria, J. C. Campuzano, T. Yokoya, 
T. Takeuchi, T. Takahashi, T. Mochiku, K. Kadowaki, P. Guptasarma and D. G. Hinks, 
Nature {\bf 392}, 157 (1998).

\bibitem{PJ} H.-B. Yang, J. D. Rameau, P. D. Johnson, T. Valla, A. Tsvelik and G. D. Gu,
Nature {\bf 456}, 77 (2008).

\bibitem{Zhou} J. Meng, G. Liu, W. Zhang, L. Zhao, H. Liu, X. Jia, D. Mu, S. Liu,
X. Dong, J. Zhang, W. Lu, G. Wang, Y. Zhou, Y. Zhu, X. Wang, Z. Xu,
C. Chen and X. J. Zhou, Nature {\bf 462}, 335 (2009).

\bibitem{Doiron} N. Doiron-Leyraud, C. Proust, D. LeBoeuf, J. Levallois, J.-B. Bonnemaison, 
R. Liang, D. A. Bonn, W. N. Hardy and  L. Taillefer, Nature {\bf 447}, 565 (2007).

\bibitem{LeBoeuf} D. LeBoeuf, N. Doiron-Leyraud, R. Daou, J.-B. Bonnemaison,
J. Levallois, N.E. Hussey, C. Proust, L. Balicas, B. Ramshaw,
R. Liang, D.A. Bonn, W.N. Hardy, S. Adachi and L. Taillefer,
Nature {\bf 450}, 533 (2007).

\bibitem{Sebastian-QCP} S. E. Sebastian N. Harrison, M. M. Altarawneh, C. H. Mielke,
R. Liang, D. A. Bonn, W. N. Hardy and G. G. Lonzarich, Proc. Natl. Acad. Sci. (online), arXiv:0910.2359.

\bibitem{Boebinger} Y. Ando, G. S. Boebinger, A. Passner, T. Kimura and K. Kishio,
Phys. Rev. Lett. {\bf 75}, 4662 (1995).

\bibitem{Proust} C. Proust, L. Taillefer, M. Sutherland, N. Doiron-Leyraud, D. LeBoeuf, J. Levallois,
M. Nardone, H. Zhang, N. Hussey, S. Adachi, R. Liang, D.A. Bonn and W. N. Hardy, unpublished.

\bibitem{Dai}
P. Dai, H. A. Mook, R. D. Hunt and F. Dogan, Phys. Rev. B {\bf 63}, 054525 (2001).

\bibitem{Hinkov} V. Hinkov, D. Haug, B. Fauque, P. Bourges, Y. Sidis, A. Ivanov,
C. Bernhard, C. T. Lin and B. Keimer, Science {\bf 319}, 597 (2008).

\bibitem{Dai2} S. Li, S. Chang, X. Yao, K. Segawa, Y.
Ando, H. A. Mook, M. R. Norman and P. Dai, unpublished.

\bibitem{Sebastian-PNAS} S. E. Sebastian, N. Harrison, P. A. Goddard, M. M. Altarawneh, 
C. H. Mielke, R. Liang, D. A. Bonn, W. N. Hardy, O. K. Andersen and G. G. Lonzarich,
arXiv:1001.5015

\bibitem{Subir} E. Demler, S. Sachdev and Y. Zhang, Phys. Rev. Lett. {\bf 87}, 067202 (2001)
and S. Sachdev, Phys. Stat. Sol. b {\bf 247}, 537 (2010).

\bibitem{LR} P. A. Lee and T. V. Ramakrishnan, Rev. Mod. Phys. {\bf 57}, 287 (1985).

\bibitem{Millis} A. J. Millis and M. R. Norman, Phys. Rev. B {\bf 76}, 220503(R) (2007).

\bibitem{Jie1} J. Lin and A. J. Millis, Phys. Rev. B {\bf 78}, 115108 (2008) and {\bf 80}, 193107 (2009).

\bibitem{Yamada} K. Yamada, C. H. Lee, K. Kurahashi, J. Wada, S. Wakimoto, S. Ueki, H. Kimura, 
Y. Endoh, S. Hosoya, G. Shirane, R. J. Birgeneau, M. Greven, M. A. Kastner and Y. J. Kim,
Phys. Rev. B {\bf 57}, 6165 (1998).

\bibitem{Audouard} A. Audouard, C. Jaudet, D. Vignolles, R. Liang, D. A. Bonn, W. N. Hardy,
L. Taillefer and C. Proust, Phys. Rev. Lett. {\bf 103}, 157003 (2009).

\bibitem{Sebastian} S. E. Sebastian, N. Harrison, E. Palm, T. P. Murphy, C. H. Mielke, 
R. Liang, D. A. Bonn, W. N. Hardy and G. G. Lonzarich, Nature {\bf 454}, 200 (2008).

\bibitem{Keimer} D. Haug, V. Hinkov, A. Suchaneck, D. S. Inosov, N. B. Christensen, 
Ch. Niedermayer, P. Bourges, Y. Sidis, J. T. Park, A. Ivanov, C. T. Lin, J. Mesot and B. Keimer,
Phys. Rev. Lett. {\bf 103}, 017001 (2009).

\bibitem{Corcoran} R. Corcoran, N. Harrison, S. M. Hayden, P. Meeson, M. Springford
and P. J. van der Wel, Phys. Rev. Lett. {\bf 72}, 701 (1994).

\bibitem{Louis} Louis Taillefer, private communication.

\bibitem{Jaudet} C. Jaudet, J. Levallois, A. Audouard, D. Vignolles, B. Vignolle, R. Liang,
D. A. Bonn, W. N. Hardy, N. E. Hussey, L. Taillefer and C. Proust, Physica B {\bf 404}, 354 (2009).

\bibitem{Rourke} P. M. C. Rourke, A. F. Bangura, C. Proust, J. Levallois, N. Doiron-Leyraud,
D. LeBoeuf, L. Taillefer, S. Adachi, M. L. Sutherland and N. E. Hussey, arXiv:0912.0175

\bibitem{Cyril} Cyril Proust, private communication.

\bibitem{Sudip} S. Chakravarty, A. Sudbo, P. W. Anderson and S. Strong,
Science {\bf 261}, 337 (1994).

\bibitem{Liang} R. Liang, D. A. Bonn and W. N. Hardy, Phys. Rev. B {\bf 73}, 180505(R) (2006).

\bibitem{Li} S. Li, Z. Yamani, H. J. Kang, K. Segawa, Y. Ando, X. Yao,
H. A. Mook and P. Dai, Phys. Rev. B {\bf 77}, 014523 (2008).

\end{thebibliography}
\end{document}